# High Thermal Conductivity of Back-End-of-Line Compatible Diamond Films

Jinwen Liu[1], Chufei Cheng[2], Feifei Tan[3], Jinquan Zhang[1], Di Lu[3, *], Bing Dai[2, *], Jiaqi Zhu[2], Runsheng Wang[1, 4], Zhe Cheng[1, 4, *]

1 School of Software & Microelectronics, Peking University, Beijing 100871, China

2 National Key Laboratory of Science and Technology on Advanced Composites in Special Environments, Harbin Institute of Technology, Harbin, 150080, China

3 Shenzhen Eversix Technology Co., Ltd., Shenzhen, Guangdong, 518083, CHN

4 School of Integrated Circuits and Beijing Advanced Innovation Center for Integrated Circuits, Peking University, Beijing 100871, China

*Authors to whom correspondence should be addressed: zhe.cheng@pku.edu.cn; daib@hit.edu.cn; ludi@eversix.com

## Abstract

Back-end-of-line (BEOL) thermal management requires electrically insulating heat-spreading dielectric that can be integrated within thermal budgets below 400°C. Here, we report polycrystalline diamond films grown directly on Si at a substrate temperature below 400°C. Two films with average thickness of ~760 and ~1000 nm were characterized by Raman spectroscopy, scanning electron microscopy (SEM), and time-domain thermoreflectance (TDTR). Raman spectra show a sharp diamond peak with minor signatures of non-diamond carbon, while SEM reveals lateral growth and large grain size. Temperature dependent TDTR measurements were performed from room temperature to 100°C. Sensitivity analysis indicates that the sensitivity of cross-plane thermal conductivity is comparative to the in-plane thermal conductivity. Accordingly, the films were analyzed using an isotropic thermal model by considering the nearly-isotropic grain structure, yielding room temperature effective thermal conductivity of ~73 and ~86 W $m^{-1}$ $K^{-1}$, respectively. These values are about two orders of magnitude higher than those of conventional dielectric materials and demonstrate the potential of diamond films grown at low temperatures as dielectric heat-spreading layers.

## Introduction

In advanced 3D stacked devices, heat is generated in highly confined regions and must be transported through multilayer stacks consisting of semiconductors, metals, dielectrics, bonding layers, and buried interfaces.[1–3] As metal interconnects continue to scale, localized Joule heating in the BEOL becomes increasingly severe, while the low thermal conductivity of conventional interlayer dielectrics and oxides impedes heat dissipation and exacerbates the resulting temperature rise.[2,4] The resulting temperature rise can degrade device performance and accelerate reliability failure.[5] These constraints motivate the development of thermally conductive dielectric materials that can be placed close to heat-generation regions without compromising electrical isolation or device operation.[6]

Diamond is well suited for this role because it combines an ultrawide bandgap of ~5.5 eV, high chemical stability, ultrahigh thermal conductivity (~2200 W $m^{-1}$ $K^{-1}$ for single crystal diamond), a relatively low dielectric constant, and electrical insulation.[5] Polycrystalline diamond can be deposited by hot filament or microwave plasma chemical vapor deposition (MPCVD), usually after substrate surface is seeded with nanodiamond particles. However, conventional diamond growth usually requires substrate temperature of 650-1000°C,[7–9] which is incompatible with completed semiconductor devices and BEOL interconnect structures. Lowering the diamond growth temperature is therefore essential if diamond is to be integrated as an heat-spreading dielectric, rather than only used as a bonded substrate, backside heat spreader, or external heat sink.[10–12]

Recent studies have shown that low-temperature diamond growth can retain useful thermal properties when nucleation density and grain evolution are properly controlled. Millán-Barba *et al*. deposited 219-420 nm nanocrystalline diamond below 450°C by microwave linear antenna plasma enhanced chemical vapor deposition system and measured surface thermal conductivity of approximately 100 W $m^{-1}$ $K^{-1}$ using scanning thermal microscopy (SThM-AFM).[13] However, their Raman spectra revealed a significant presence of non-diamond carbon phase within these films. Furthermore, their thermal measurements heavily depended on a calibration reference, assuming a constant thermal conductivity of 200 W $m^{-1}$ $K^{-1}$ for the Si substrate. Lundh *et al*. extended low-temperature diamond to Al-rich AlGaN HEMTs by growing ~250 nm thick nanocrystalline diamond heat spreader at 500°C, which had a thermal conductivity of 45±25 W $m^{-1}$ $K^{-1}$ sdf.[14] More recently, Tzeng *et al*. reported ~100 nm MPCVD diamond films grown under 450°C, with effective thermal conductivity approaching or exceeding 300 W $m^{-1}$ $K^{-1}$ under optimized seeding and plasma conditions.[15] Nevertheless, reliable extraction of the thermal conductivity of such ultrathin and highly conductive films remains challenging because their thermal resistance is small relative to the thermal boundary resistances, resulting negligible TDTR sensitivity to the film thermal conductivity and strong parameter correlation with the thermal boundary conductance. Notably, Sood and Cahill were unable to obtain reliable through-plane thermal conductivity data even for a 0.5 μm thick diamond film because of insufficient measurement sensitivity.[16] Considering that the films investigated by Tzeng *et al*. were substantially thin, rigorous sensitivity analysis is essential for evaluating the reliability of the extracted values.

Malakoutian *et al*. demonstrated polycrystalline diamond growth at 400°C by optimizing the $H_2/CH_4/O_2$ gas chemistry and staged growth process, obtaining near-isotropic grains and a thermal conductivity of ~300 W $m^{-1}$ $K^{-1}$.[17] Their subsequent study of the 300-400°C growth window further showed that introducing oxygen is critical for maintaining a pronounced $sp^3$ Raman peak and suppressing $sp^2$ carbon phase at low temperature.[18] By promoting lateral grain coarsening and suppressing smaller grains during the early growth stage, diamond films spanning approximately 0.3-25 μm were reported with thermal conductivity between about 300 and 1800 W $m^{-1}$ $K^{-1}$ under the temperature of 400-650°C.[19] When these physical and thermal properties were incorporated into simulation models, the temperature reduces ~20% and ~50% in a flip-chip structure and monolithic three-dimensional accelerator, respectively.[19]

For polycrystalline diamond films with BEOL compatible thickness, thermal conductivity should not be evaluated by comparison with bulk diamond alone. In films from several hundred of nanometers to a few micrometers, phonon transport is governed by grain size, boundary disorder, columnar growth, film thickness, impurities, and $sp^2$ carbon.[17,20] Angadi *et al*. showed that nanocrystalline diamond with 3-5 nm grains exhibits thermal conductivity only up to about 12 W $m^{-1}$ $K^{-1}$, illustrating the strong effect of grain boundary on phonon transport.[21] Anaya *et al.* further demonstrated that the in-plane thermal conductivity of ultrathin nanocrystalline diamond can be tuned over a broad range, by controlling grain size and quality.[22] In suspended polycrystalline diamond membranes, Sood *et al.* used TDTR to show that films near 1μm can exhibit markedly different in-plane and cross-plane thermal conductivity, and that the extracted values depend on film thickness and on whether the growth side or nucleation side is measured.[16] Under conventional CVD growth conditions, reducing the substrate temperature generally promotes the incorporation of non-diamond phase and defects, which enhance phonon scattering and consequently tend to lower the thermal conductivity of diamond films. These reports provide a necessary reference frame for assessing low-temperature diamond films.

Besides, recent studies have reported remarkably high thermal conductivities in diamond films grown at reduced temperatures, although substantial discrepancies remain among the reported values. A major reason for these discrepancies is the difficulty of accurately measuring thermal transport in such ultrathin films, whose small intrinsic thermal resistance can lead to low measurement sensitivity. To help reconcile these discrepancies, we investigate the thermal conductivity of diamond films grown below 400 °C using TDTR with sensitivity analysis.

In this work, we report the thermal conductivity of polycrystalline diamond films directly grown on Si at a substrate temperature lower than 400℃. The crystal quality and morphology of the films were characterized by Raman spectroscopy and SEM. TDTR sensitivity analysis was performed to evaluate the response of the measured signal to the film thermal properties. The thermal conductivity of the films was then determined by TDTR.

**Results and Discussion**

The samples were grown on Si substrates at a substrate temperature of <400℃ via microwave plasma chemical vapor deposition (MPCVD). The growth process utilized the MPDF-6K-L type equipment produced by Shenzhen Eversix Technology Co., Ltd. The detailed growth procedure, including the seeding protocol, gas mixture, and growth-window optimization, will be reported elsewhere. Briefly, the growth condition used here was selected to enable diamond deposition compatible to BEOL temperature window.

The growth surface of the diamond exhibited obvious roughness, leading to low optical reflectance and unreliable TDTR signals. Therefore, the diamond films were mechanically released from the Si substrate using a Scotch tape and transferred onto a Kapton tape on a glass substrate (Fig. 1a). The nucleation side of the diamond film, originally adjacent to the Si substrate and having nanometer-scale roughness, was exposed for subsequent Aluminum transducer deposition and TDTR measurement. The thickness of the two films used for TDTR measurements were determined by stylus profilometry to be ~760 and ~1000 nm.

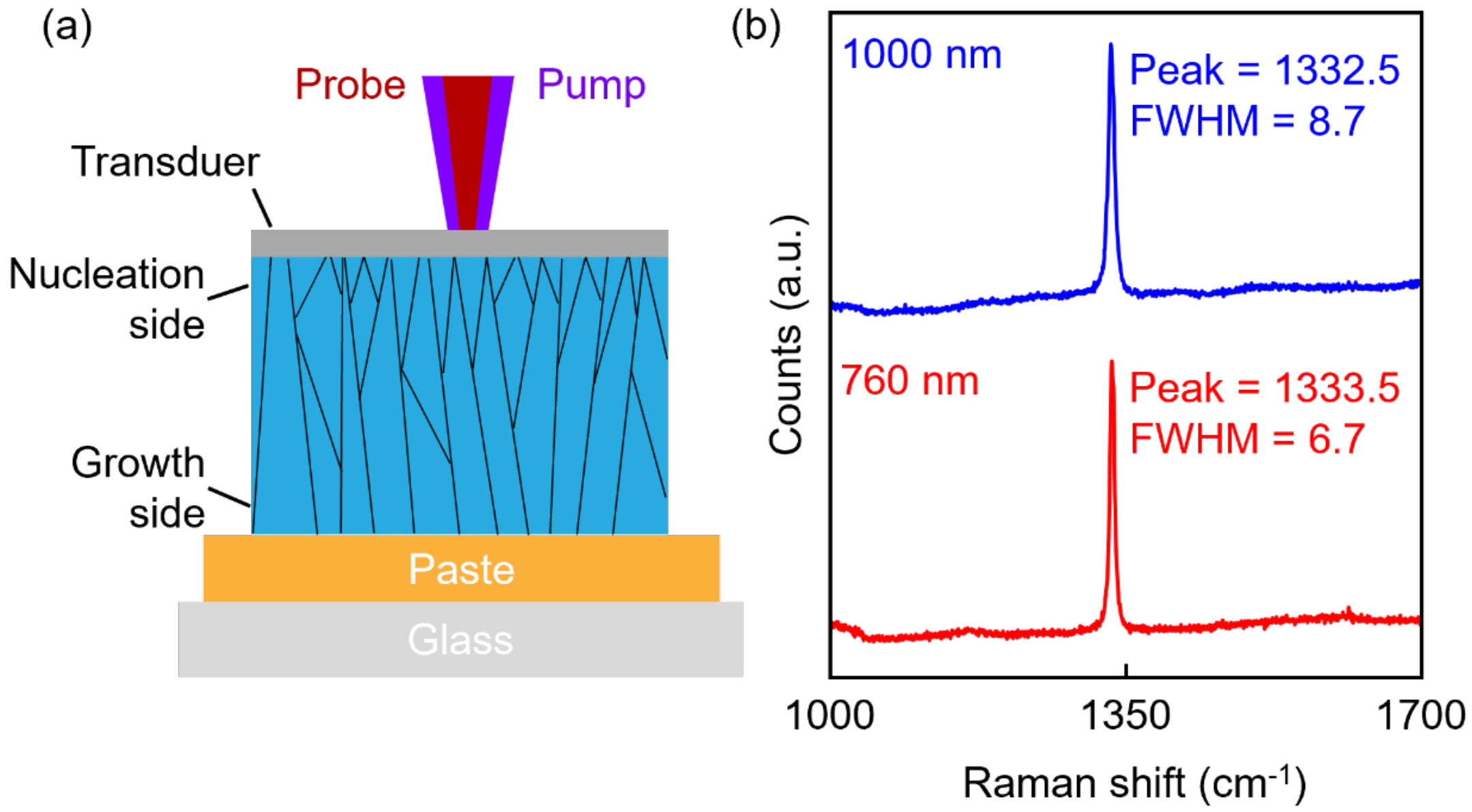

**Figure 1. TDTR measurement configuration and Raman characterization**. (a) Schematic of the TDTR measurement geometry after mechanical transfer of the diamond film and deposition of the Al transducer on the nucleation side. (b) Raman spectra of the diamond films.

As shown in Fig. 1b, Raman results were used to verify the diamond phase and the crystalline quality of our samples. Both spectra show a clear diamond peak near 1332 $cm^{-1}$, confirming the formation of $sp^3$ bonds on Si substate under low temperature. The peak positions are 1332.5 $cm^{-1}$ and 1333.5 $cm^{-1}$ for the 1000 and 760 nm films, respectively., with corresponding full widths at half maximum (FWHM) of 8.7 $cm^{-1}$ and 6.7 $cm^{-1}$. Besides, our samples exhibited minimal non-diamond phase. Because Raman peaks width and $sp^2$ bonds are closely linked to the crystal quality,[17] these spectra indicate that the present low-temperature films retain a relatively high diamond phase purity.

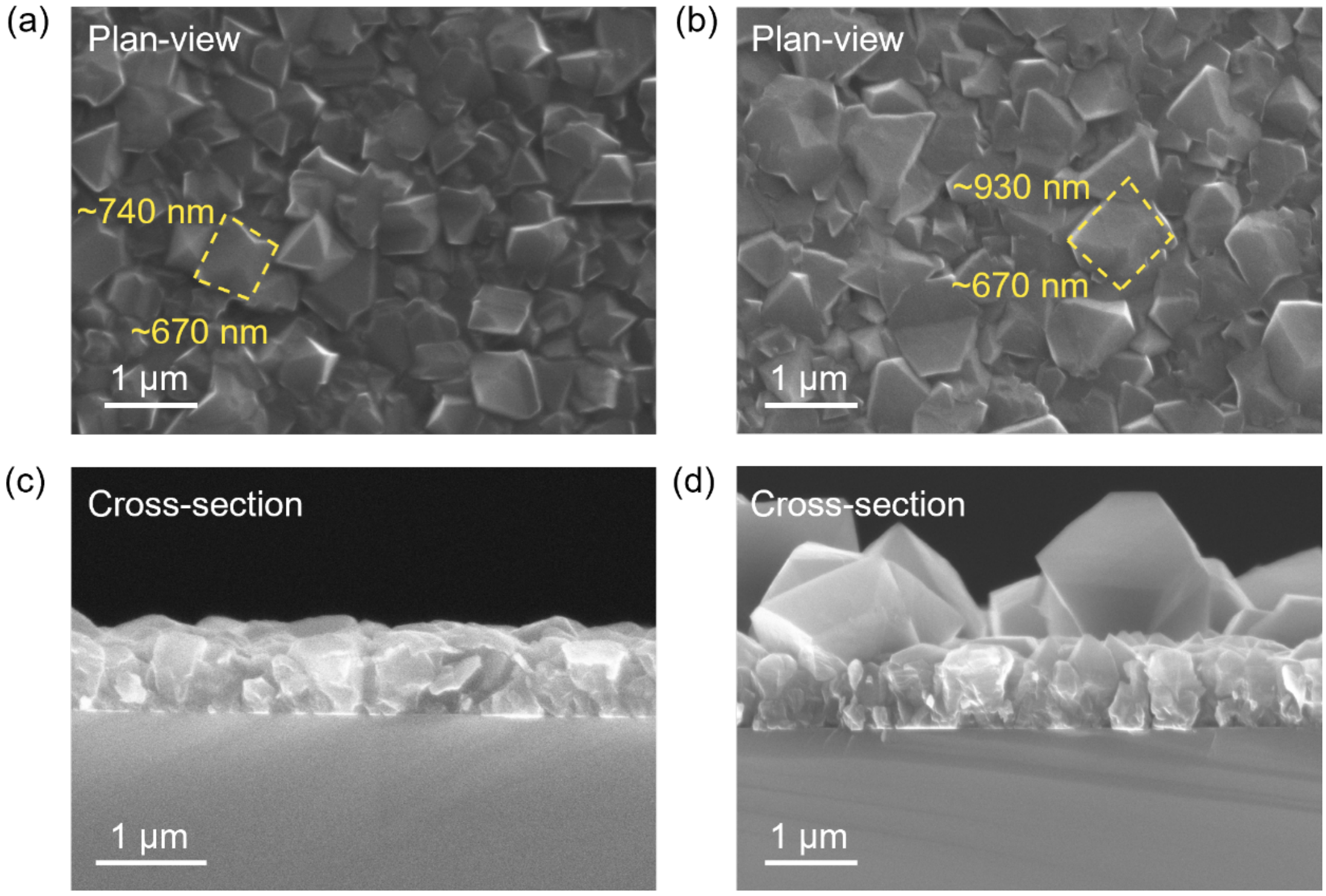

**Figure 2. Representative morphology of the diamond films.** (a, b) Plan-view and (c, d) cross-sectional SEM images acquired from separate specimens grown under the same conditions as the films used for TDTR measurements.

Representative plane-view and cross sectional SEM images of diamond films grown under the same conditions are shown in Fig. 2. These results demonstrate the large grain size under the low-temperature growth conditions. Such grain development is important for heat spreading because grain-boundary scattering is a primary factor limiting phonon transport in polycrystalline diamond films.[22]

The thermal conductivity was measured by a two-color TDTR system. The pump and probe wavelengths were 400 and 800 nm, respectively. The pump beam was modulated at 10.08 MHz to periodically heat the Al transducer, while the probe beam monitored the thermoreflectance response. The $1/e^2$ radius of the pump and probe beams were 4.38 and 2.7 μm, respectively. Temperature dependent measurements were performed on a temperature-controlled stage from room temperature to 100°C. More details about our TDTR can be found in Ref.[23,24]

Fig. 3a shows the temperature dependent effective thermal conductivity of the 760 and 1000 nm diamond films. At room temperature, the effective thermal conductivity values are ~73 W $m^{-1}$ $K^{-1}$ and ~86 W $m^{-1}$ $K^{-1}$, respectively. The increase in thermal conductivity with film thickness is consistent with the grain development, since thicker polycrystalline diamond films generally contain a larger fraction of grains and therefore reduce boundary scattering.[22] Both films exhibit an empirical exponential decay temperature dependence relationship.[25]

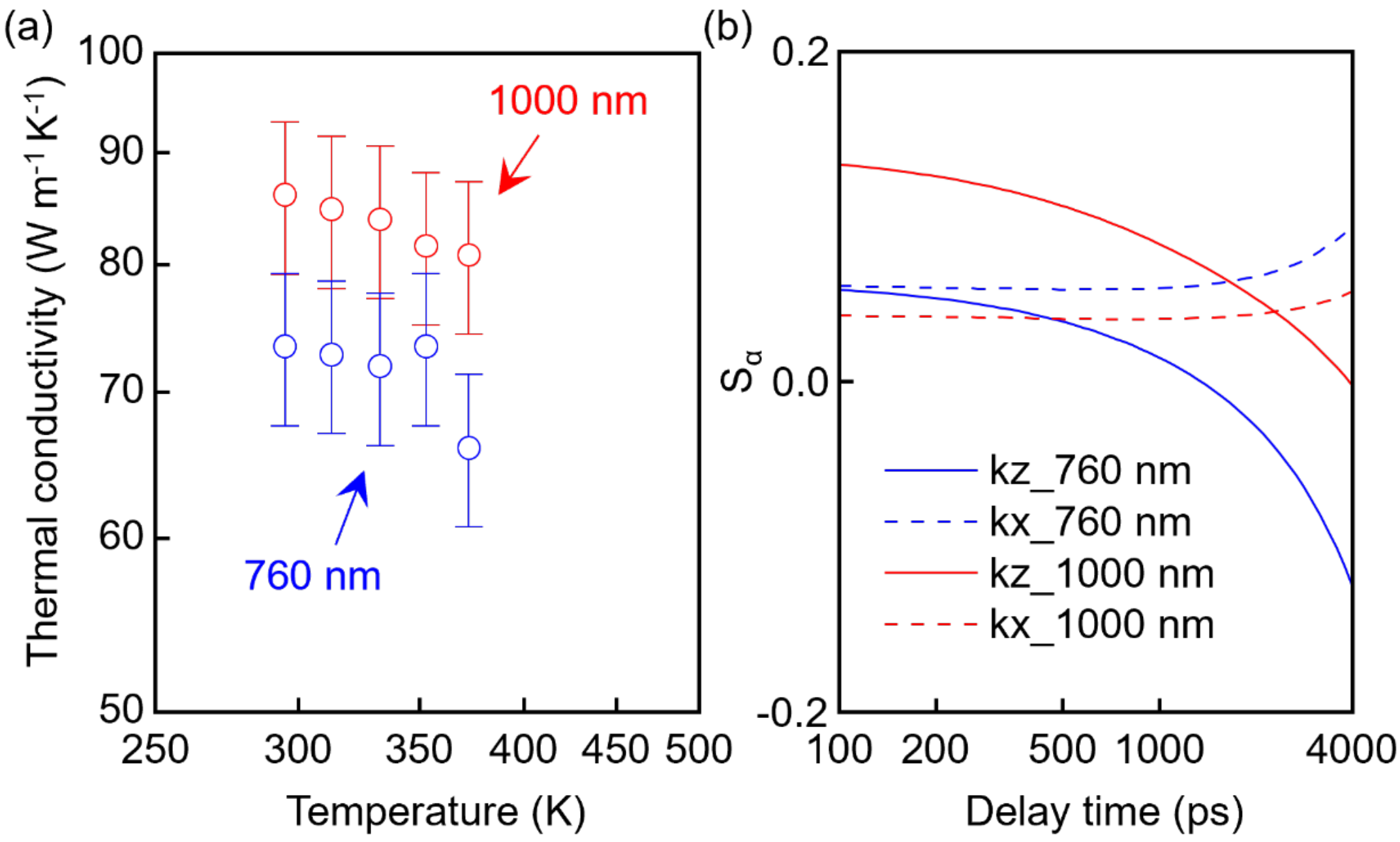

**Figure 3. Thermal conductivity of diamond films.** (a) Effective thermal conductivity extracted from TDTR measurement of the 760 nm and 1000 nm samples. (b) Sensitivity coefficients of in-plane and out-of-plane thermal conductivity, calculated using $k_{Al}$ = 200 W m$^{-1}$ K$^{-1}$ which was derived from four probe measurements and the Wiedemann-Franz law, $k_{diamond}$ = 73 W m$^{-1}$ K$^{-1}$, $TBC_{Al/diamond}$ = 77.8 MW m$^{-2}$ K$^{-1}$, $th_{Al}$ = 78 nm, $th_{diamond}$ = 760 nm, $TBC_{diamond/tape}$ = 18 MW m$^{-2}$ K$^{-1}$, $k_{tape}$ = 0.15 W m$^{-1}$ K$^{-1}$, $k_{diamond}$ = 85 W m$^{-1}$ K$^{-1}$, $TBC_{Al/diamond}$ = 52.5 MW m$^{-2}$ K$^{-1}$, $th_{Al}$ = 83 nm, $th_{diamond}$ = 1000 nm, $TBC_{diamond/tape}$ = 20 MW m$^{-2}$ K$^{-1}$, $k_{tape}$ = 0.15 W m$^{-1}$ K$^{-1}$.

The present TDTR measurements were performed from the nucleation side. In polycrystalline diamond, the nucleation side usually contains smaller grains and a higher density of grain boundaries. Sood *et al*. showed that measurements from the growth side and nucleation side can give different thermal conductivity because TDTR is more sensitive to the region close to the heat source.[16] Therefore, nucleation-side measurements may contribute to a lower value than the growth side. As shown in Fig. 3b, the TDTR ratio exhibits comparable sensitivity of both the in-plane and out-of-plane thermal conductivity. The data were therefore analyzed using an isotropic thermal model in which $\kappa_r$ and $\kappa_z$ were constrained to a common value.

Table 1 summarizes the thermal conductivity values reported for thin diamond films grown over a wide range of substrate temperature. The substantial variation among the reported values, even for films with comparable thickness, highlights the combined influence of growth conditions, microstructure, and measurement geometry.

Table 1. Comparison of room temperature thermal conductivity values reported for diamond films.

| | Growth temperature (°C) | Film thickness (nm) | Method | $\kappa$ (W m$^{-1}$ K$^{-1}$) | Raman FWHM (cm$^{-1}$) |
|---|---|---|---|---|---|
| This work | <400 | 760 | TDTR | ~73 (Nucleation side, $\kappa_{eff}$) | 6.7 |
| This work | <400 | 1000 | TDTR | ~86 (Nucleation side, $\kappa_{eff}$) | 8.7 |
| Malakoutian[17] | 400 | 798 | TTR | ~300 ($\kappa_z$) | 8.82 |
| Tzeng[15] | 450 | 130.5 | TDTR | ~300 ($\kappa_z$) | ~13.0 |
| Anaya[22] | 825 | 980 | Raman | 218 ($\kappa_r$) | … |
| Anaya[22] | 825 | 880 | Raman | 198 ($\kappa_r$) | … |

| Anaya[22] | 750 | 470 | Raman | 115 ($\kappa_r$) | … |
|---|---|---|---|---|---|
| Anaya[22] | 750 | 680 | Raman | 133 ($\kappa_r$) | … |
| Sood[16] | 750 | 1000 | TDTR | 77 ($\kappa_r$)/210 ($\kappa_z$) | … |
| Sood[16] | 750 | 1000 | TDTR | 46 ($\kappa_r$)/89 ($\kappa_z$) (Nucleation side) | … |
| Yates[8] | 750 | 1000 | TDTR | 103 ($\kappa_r$)/175 ($\kappa_z$) | … |
| Lundh[14] | 500 | 250 | TDTR | 45 ($\kappa_z$) | ~10.4 |
| Hao[26] | 700 | 150 | TDTR | 200 ($\kappa_z$) | ~11.6 |
| Wu[27] | 800 | 1500 | TDTR | ~314 ($\kappa_z$) | ~40 |
| Cheng[7] | 750 | 2300 | TDTR | 363 ($\kappa_z$) | … |

Sensitivity analysis is essential for evaluating the reliability of thermal parameters extracted from TDTR measurements, particularly for submicrometer, highly thermally conductive films. The TDTR sensitivity depends not only on the intrinsic properties of the film but also on the modulation frequency and the laser spot size. If the sensitivity is low, similar agreement between the experimental data and the thermal model may be obtained using different combinations of thermal properties. A satisfactory model fit alone therefore does not necessarily demonstrate that a reasonable parameter has been determined. Therefore, the measurement configuration should be carefully designed to ensure sufficient sensitivity to the parameter of interest.

**Summary**

Polycrystalline diamond films were grown on Si at ~400°C. Raman spectroscopy reveals narrow diamond peaks with FWHM values of 6.7 and 8.7 $cm^{-1}$ and only minor non-diamond phase, confirming the relatively high crystalline quality of the films. Sensitivity analysis shows that the measured TDTR ratio is influenced by both in-plane and out-of-plane heat transport. The data were therefore analyzed using an isotropic TDTR model, where $\kappa_r = \kappa_z = \kappa_{eff}$. Despite being grown below 400 °C, the films exhibit effective thermal conductivities (73 W $m^{-1}$ $K^{-1}$ for 760 nm and 86 W $m^{-1}$ $K^{-1}$ for 1000 nm) substantially higher than those of conventional BEOL dielectric materials, supporting their potential as electrically insulating heat-spreading layers.

**Conflict of Interest**

The authors declare that they have no conflict of interest.

**Acknowledgements**

The authors acknowledge the financial support from the National Key Research and Development Program of China (Grant No.2024YFA1207901) and the National Natural Science Foundation of China (NSFC) (Grant Nos. 62574007, T2550270). This work was supported by the MIND project (MINDXZ202602).

**Data availability**

The data that support the findings of this study are available from the corresponding authors upon reasonable request.